\documentclass{article}

\usepackage{amsthm,amsmath,amsfonts,amssymb}
\usepackage{natbib}

\usepackage[latin1]{inputenc}
\usepackage[english]{babel} % hyphenation patterns for American English
\usepackage{subfigure}
\usepackage{setspace}
\usepackage{tabularx} % para larguras das colunas das tabelas
\usepackage{ctable} % for \toprule, \midrule and \bottomrule
\usepackage{enumerate}
\usepackage{epsfig}
\usepackage{graphics}
\usepackage{graphicx}
\usepackage{color}
\usepackage{slashbox} 
\usepackage{booktabs} 
\usepackage{longtable}
\usepackage{url}

\usepackage[short,12hr]{datetime}
\usepackage{fullpage}
\usepackage{srcltx}
\usepackage{flushend}
\usepackage{mathptmx}

\allowdisplaybreaks[4] % para evitar espaços entre fórmulas

\title{Bootstrap Bartlett correction in inflated beta regression}

\author{
La\'is H. Loose\thanks{L.~H.~Loose 
is with the
Departamento de Estat\'istica,
Universidade Federal do Rio Grande do Norte, RN, Brazil,
E-mail: laisloose@gmail.com}
\and
F\'abio M. Bayer\thanks{F.~M.~Bayer 
is with the
Departamento de Estat\'istica and LACESM,
Universidade Federal de Santa Maria, RS, Brazil,
E-mail: bayer@ufsm.br}
\and
Tarciana L. Pereira\thanks{T.~L.~Pereira  is with the
Departamento de Estat\'istica,
Universidade Federal da Paraíba, PB, Brazil,
E-mail: tarciana@de.ufpb.br}
}
\date{}

\begin{document}

\maketitle

\doublespacing

\abstract{
\noindent
The inflated beta regression model aims to enable the modeling of responses in the intervals $(0,1]$, $[0,1)$ or $[0,1]$. 
In this model, hypothesis testing is often performed based on the likelihood ratio statistic.
The critical values are obtained from asymptotic approximations, 
which may lead to distortions of size in small samples.
In this sense, 
this paper proposes the bootstrap Bartlett correction to the statistic of likelihood ratio in the inflated beta regression model. The proposed adjustment only requires a simple Monte Carlo simulation. Through extensive Monte Carlo simulations the finite sample performance (size and power) of the proposed corrected test 
is compared to the usual likelihood ratio test and the Skovgaard adjustment already proposed in the literature.
The numerical results evidence that inference based on the proposed correction is much more reliable than that based on the usual likelihood ratio statistics and the Skovgaard adjustment.
At the end of the work, an application to real data is also presented.


\section{Introduction} 

The beta regression model proposed by~\cite{Ferrari2004}
is appropriate when the dependent variable assumes values in the standard unit interval $(0 , 1)$, such as rates, proportions or indexes. 
It is assume that the response follows a beta law with 
constant precision parameter and mean parameter modeled by a regression structure. This regression structure is similar to the generalized linear model (GLM)~\citep{McCullagh1989}. The mean response is related to a linear predictor through a link function and the linear predictor involves known covariates and unknown regression parameters \citep{Ospina2006, Bayer2013}. 
In \cite{Parker2014} the authors present a discussion about the origins of beta regression models.

In rates and proportions data, zeros and/or ones values can often be observed. For example, when the mortality rate for a given disease, child labor rate, proportion of hospital admissions for certain cause, among other situations, are to be evaluated. In such cases the seminal model proposed in~\cite{Ferrari2004} is not suitable. The log-likelihood function of the beta regression model becomes non-limited, and it's not possible to assume that data come from an absolutely continuous distribution. For these cases, \cite{Ospina2012} propose the inflated beta regression model, based on mixture of beta and bernoulli degenerate at zero and/or one distributions. It is important to mention that a degenerate distribution is the probability distribution of a discrete random variable that assumes probability 1, to a single point \citep{sundarapandian2009}. These inflated distributions allow users to model data that assume values in $(0,1]$, $[0,1)$ or $[0,1]$ \citep{Ospina2010}. 
In this work it will be addressed the model of inflated beta regression in zero or one. 

The probability density function of the inflated beta distribution at zero or one has three parameters: conditional mean ($\mu_t$), precision ($\phi_t$) and the mixture parameter ($\alpha_t$). The latter determines the probability that the dependent variable is equal to one of the limits of the unit interval. In the inflated beta regression model, each one of these parameteres is assumed to be variable along the observations, being modeled using regression structures that involve link functions, covariates and unknown parameters. The presence of regression structures for the three parameters that index the inflated beta density makes the problem of inferences in small samples more severe, given the large number of parameters to be estimated. 

The estimation of the inflated beta regression model's parameters is based on maximum likelihood estimation (MLE), in which the inferential procedures are similar to GLM. After the point estimation, another important aspect in the modeling are the hypothesis testing on the parameters of the model. One of the usual test statistics to perform hypothesis testing is the likelihood ratio (${{\rm{LR}}}$)~\citep{Person1928}. This is an approximate test and is characterized by the use of critical values from approximations that are valid in large samples. However, these asymptotic approximations can be poor in small samples, resulting in considerable distortion of the probability of type I error (size) of the tests. Inferential improvements in small samples may be achieved by analytical or numerical/computational adjustments. Two important works on hypotheses testing and finite corrections to asymptotic tests are, respectively, \cite{Buse1982} and \cite{Cribari1996}. 

Several studies have been developed to improve the performance of the likelihood ratio test in small samples. Among the proposals for inferencial improvement stands out the Bartlett correction \citep{Bartlett1937}, in which its analytical derivation involves cumulants and mixed cumulants up to fourth order of the log-likelihood function. In \cite{cysneiros2006}, the Bartlett correction is presented in non-linear models of the exponential family. For improvements of the heteroscedasticity test in the normal linear regression model, \cite{cysneiros2004} use this correction. In \cite{melob2009}, the Bartlett correction is derived from the class of linear mixed models. Also, in \cite{Bayer2013}, the Bartlett correction in the beta regression model with constant dispersion is considered. However, the derivation of the Bartlett correction can be costly, or even impossible to obtain \citep{Pinheiro2011, Bayer2013}, especially when the parameters are not orthogonal, as in the inflated beta regression model. 

Another alternative is the Skovgaard adjustment \citep{skovgard2001}. Some recent papers consider this adjustment were developed in the class of nonlinear models of the exponential family \citep{cysneiros2008}, in a new class of models for proportions \citep{lemonte2009}, in the beta regression model with variable dispersion \citep{Pinheiro2011} and for the model of inflated beta regression \citep{Pereira2012}. Despite the Skovgaard adjustment being less analytical costly than the Bartlett correction, it still requires second-order derivatives of the log-likelihood function, being that a limitation primarily to inferential improvements in applied works. 

With the same objective of the Skovgaard and Bartlett adjustments, which is to improve the approximation of the chi-squared distribution to the exact null distribution of the likelihood ratio statistic in small samples, it can be considered the bootstrap Bartlett correction \citep{Rocke1989}. In this second-order correction, the Bartlett correction factor \citep{Lawley1956} is determined by the bootstrap method \citep{Efron1979}. The bootstrap Bartlett correction becomes a good numerical alternative to analytical determination of the Bartlett correction factor, requiring only the use of a simple Monte Carlo simulation. The bootstrap Bartlett correction still has computational advantages over the usual bootstrap procedure for the determination of exact quantiles for the null distribution of the test statistic. While the usual bootstrap method requires a large number of resamples (usually above 1000), the numerical Bartlett correction requires a smaller number of bootstrap iterations (around 200 resamples)~\citep{Bayer2013}. Despite extensive advantages in using the bootstrap Bartlett correction versus other analytical and numerical approaches, this approach is rarely explored in the literature. One of the few studies that consider the bootstrap Bartlett correction was developed by \cite{Bayer2013}, evidencing similar results between the analytical and bootstrap Bartlett corrections. 

In order to improve the inferences in small samples in the inflated beta regression model, this work proposes the bootstrap Bartlett correction to the likelihood ratio statistic. The performance in small samples of the proposed test statistic is compared with the Skovgaard adjustment \citep{Pereira2012} and the usual likelihood ratio statistics, via Monte Carlo simulations. The approximations of statistics' distributions by chi-squared distribution in samples of finite size are evaluated, and the influences of these approximations on the performance of hypothesis testing are verified, in terms of size and power of the tests. 

This paper is organized as following. Section~\ref{s:modelo} introduces the inflated beta regression model at zero or one, as well as link functions, log-likelihood function and inferential details. In Section~\ref{S:TesteRV}, the likelihood ratio test for the inflated beta regression model, the proposed bootstrap Bartlett correction and Skovgaard adjustment for small samples are presented. Section~\ref{S:resultados} describes the experiment of Monte Carlo simulation for finite samples and presents the numerical results and its discussion. In Section~\ref{s:aplicacao}, an application to real data is presented and discussed. Finally, Section~\ref{s:conclu} presents the conclusions. 

\section{Zero-or-one inflated beta regression model}\label{s:modelo} 

The beta regression model proposed in \cite{Ferrari2004} is based on a reparametrization of the beta density, indexed by parameters of mean $\mu$ and precision $\phi$. The parameter $\phi$ is considered constant and $\mu$ is modeled by a regression structure. The beta density is given as follows: 
\begin{align}\label{E:densidadeBeta} 
f(y;\mu,\phi)=\frac{\Gamma(\phi)}{\Gamma(\mu \phi)\Gamma((1-\mu)\phi)}y^{\mu\phi-1}(1-y)^{(1-\mu)\phi-1}, \quad 0<y<1,
\end{align} 
where $0 < \mu < 1$, $\phi>0$ and $\Gamma(\cdot)$ is the gamma function, i.e. $\Gamma(u)=\int_0^\infty t^{u-1} e^{-t}\rm{d}t$. Thus, if $y$ is a random variable with density given by Equation~\eqref{E:densidadeBeta}, we have: \begin{align*} 
{\rm{E}}(y)=& \mu, \\ {\rm{Var}}(y)=& \mu(1-\mu)/(1+\phi). 
\end{align*} 

For the inflated beta regression model a distribution for the dependent variable in which its density involves three parameters is assumed. Let $y_1, \ldots, y_n$ independent random variables, in which $y_t$, $t = 1, \ldots, n$, have inflated beta distribution at the point $c$ ($c=0$ or $c=1$), for which the density is given by \citep{Pereira2012}:
 \begin{align}\label{E:densidadeinflacionada} 
bi_c (y_t; \alpha_t, \mu_t, \phi_t) = \lbrace \alpha_t^{\mathbf{l}_{\lbrace c \rbrace} (y_t) } (1- \alpha_t )^{1-{\mathbf{l}_ {\lbrace c \rbrace} (y_t)}} \rbrace \lbrace f(y_t; \mu_t, \phi_t) ^{1-\mathbf{l}_{ \lbrace c \rbrace } (y_t) }\rbrace, 
\end{align} 
in which $\mathbf{l}_{\lbrace c \rbrace} (y_t)$ is an indicator function that assumes value $1$ if $y_t = c$ and $0$ otherwise, $0 < \alpha_t < 1$ is the mixture parameter of the distribution specified by $\alpha_t = {\rm{Pr}}(y_t = c)$, ($c=0$ or $c=1$), $0<\mu_t < 1$ is the mean of $y_t$ conditional on $y_t \in (0, 1)$, $\phi_t > 0$ is the precision parameter and $ f(y_t; \mu_t, \phi_t)$ is the beta density function given in Equation~\eqref{E:densidadeBeta}. If $c=1$, the function given in Equation~\eqref{E:densidadeinflacionada} is the density of a random variable with inflated beta distribution at one,  $y\sim {\rm BEOI}(\alpha,\mu,\phi)$. On the other hand, if $c=0$, we have an inflated beta distribution at zero, $y\sim {\rm BEZI}(\alpha,\mu,\phi)$. For $y_t$ with inflated beta distribution in $c$, where $c = 0$ or $c = 1$, expectancy and variance $y_t$ are given by~\citep{Ospina2010, Ospina2012}: 
\begin{align*} 
{\rm E}(y_t) = &\, \alpha_tc+(1-\alpha_t) \mu_t ,\\ {\rm{Var}}(y_t) = &\, (1-\alpha_t)\mu_t(1-\mu_t)/(\phi_t +1)+ \alpha_t(1-\alpha_t)(c-\mu_t)^2. 
\end{align*} 

Thus, in the zero-or-one inflated beta regression model with varying dispersion, we have the following relations~\citep{Ospina2012, Pereira2012}: 
\begin{align*} 
g(\mu_t)= &\, \sum_{i=1}^m x_{it}\beta_i=\eta_t,\\ 
b(\phi_t)= &\, \sum_{i=1}^p s_{it}\lambda_i=\kappa_t,\\ 
h(\alpha_t)= &\, \sum_{i=1}^M z_{it}\gamma_i=\zeta_t, 
\end{align*} 
with $t=1,\ldots, n$, in which $\beta=(\beta_1, \ldots , \beta_m)^{\top}$, $\lambda=(\lambda_1, \ldots , \lambda_p)^{\top}$ and $\gamma=(\gamma_1, \ldots , \gamma_M)^{\top}$ are vectors with unknown parameters, where $\beta \in \mathbb{R}^m$, $\lambda \in \mathbb{R}^p$ and $\gamma \in \mathbb{R}^M$, $x_{1t}, \ldots , x_{mt}$, $s_{1t}, \ldots , s_{pt}$ and $z_{1t}, \ldots , z_{Mt}$ represent the fixed and known covariates $(m+p+M<n)$, $g(\cdot)$, $b(\cdot)$ and $h(\cdot)$ are strictly monotonic and twice differentiable link functions, such that $g: (0,1) \rightarrow \mathbb{R}$, $b: (0,\infty) \rightarrow \mathbb{R}$ and $h: (0,1) \rightarrow \mathbb{R}$ \citep{Pereira2012}. 
Different link functions can be used: the logit, $g(\mu)=\log[\mu/(1-\mu)]$; 
the probit, $g(\mu)=\Phi^{-1}(\mu)$, in which $\Phi(\cdot)$ is the normal distribution function;
the complementary log-log, $g(\mu)=\log[-\log(1-\mu)]$;
the log-log, $g(\mu)=\log[-\log(\mu)]$;
and the Cauchy, $g(\mu)=\tan(\pi (\mu - 0.5))$;
both for $\mu$ and $\alpha$. 
For the structure of $\phi$, we have: the logarithmic function,  $b(\phi)=\log(\phi)$;
and the square root, $b(\phi)=\sqrt{\phi}$. 
For details on link functions see \cite{McCullagh1989} and \cite{Koenker2009}. 

To obtain the maximum likelihood estimators of the parametric vector $\theta= ( \beta^{\top}, \lambda^{\top}, \gamma^{\top})^{\top}$ is necessary to maximize the logarithm of the likelihood function. The log-likelihood function for $\theta= ( \beta^{\top}, \lambda^{\top}, \gamma^{\top})^{\top}$ can be written in the following way~\citep{Pereira2012}:
 \begin{align}\label{E:logverossimilhanca}
  \ell(\theta)= \lbrace (y^c- \mu^c)^{\top} \alpha^* + a^{\top} + [(y^*-\mu^*)^{\top}(\Phi \mathcal{M} - \mathcal{J})+ (y^{\dagger}-\mu^{\dagger})^{\top}(\Phi-2\mathcal{J}) + b^{\top}]H \rbrace \iota,
\end{align} 
 in which $y^c= (y^c_1, \ldots , y^c_n)^{\top}$, $y^*= (y^*_1, \ldots , y^*_n)^{\top}$, $y^{\dagger}= (y^{\dagger}_1, \ldots , y^{\dagger}_n)^{\top}$, $\mu^c= (\mu^c_1, \ldots , \mu^c_n)^{\top}$, $\mu^*= (\mu^*_1, \ldots , \mu^*_n)^{\top}$, $\mu^{\dagger}= (\mu^{\dagger}_1, \ldots , \mu^{\dagger}_n)^{\top}$, $a= (a_1, \ldots , a_n)^{\top}$, $b= (b_1, \ldots , b_n)^{\top}$, $a_t= \log (1-\alpha_t)+\mu_t^c \alpha_t^*$ and $b_t= \log \Gamma(\phi_t)- \log \Gamma(\mu_t \phi_t)- \log \Gamma((1-\mu_t)\phi_t)+(\mu_t \phi_t -1) \mu_t^* + (\phi_t-2)\mu_t^{\dagger}$. 
Moreover $\alpha^* = \rm{diag}\lbrace \alpha_1^*, \ldots , \alpha_n^* \rbrace$, $\mathcal{M} = \rm{diag}\lbrace \mu_1, \ldots , \mu_n \rbrace$, $ H = \rm{diag}\lbrace 1- y_1^c, \ldots , 1-y_n^c \rbrace$ and $ \Phi = \rm{diag}\lbrace \phi_1, \ldots , \phi_n \rbrace$ are diagonal matrices $n \times n$, $\mathcal{J}$ is the identity matrix $n \times n$ and $\iota$ is the column vector $n$-dimensional of 1s, where $\alpha^*_t=\log (\alpha_t/(1-\alpha_t))$, 
\begin{equation*} 
y_t^c= \left\{\begin{array}{ll} 1, &y_t=c,\\ 0,& y_t \in (0,1), \end{array} \right. ,\quad \quad y_t^{*}= \left\{\begin{array}{ll} \log\left(\frac{y_t}{1-y_t}\right), &y_t \in (0,1),\\ 0,& y_t=c, \end{array} \right. 
\end{equation*}
\begin{equation*}
\textrm{and} \quad y_t^{\dagger}= \left\{\begin{array}{ll} \log(1-y_t), &y_t \in (0,1),\\ 0,& y_t=c. \end{array} \right.
\end{equation*} 

For details on inferences in large samples and matrix expressions of the score vector and the Fisher information matrix, see \cite{Ospina2012} and \cite{Pereira2012}. It is noteworthy that the maximum likelihood estimators do not have closed form, being necessary the use of iterative numerical methods for maximizing the log-likelihood function, such as Newton method or quasi-Newton methods such as BFGS \citep{press}. 

The inflated beta regression model is part of the class of
generalized additive models for location, scale and shape (GAMLSS) \citep{rigby2005}. 
Thus, adjustments of inflated beta regression models considered in this work are made using the {\tt gamlss} package \citep{Stasinopoulos2007} available in the environment {\tt R}~\citep{R2012}. 
The log-likelihood maximizations
were carried out using the RS algorithm, which is a generalization of the algorithm used
by \cite{rigby1996b,rigby1996a} for fitting mean and dispersion additive
models (MADAM) \citep{Stasinopoulos2008}. 
This algorithm is well suited for situations in which the parameters are orthogonal, 
%(since the expected values of the log-likelihood cross derivatives equal zero),
and it does not require accurate starting values for the parameters to achieve convergence (the
default starting values, often constants, are usually adequate) and handles large data sets
quite efficiently \citep{Stasinopoulos2008}.

\section{Likelihood ratio test and small sample corrections}\label{S:TesteRV} 

Let $y_1, \ldots , y_n$ be independent random variables and assume that each $y_t$, $t=1, \ldots, n$, has density function given by \eqref{E:densidadeinflacionada}. Additionally, let $\theta = (\beta^{\top}, \lambda^{\top}, \gamma^{\top})^{\top}$ be the vector of unknown parameters that index the inflated beta regression model at zero or one. Consider the parameters vector $\theta=(\nu^{\top}, \tau^{\top})^{\top}$, wherein $\nu = (\nu_1, \ldots, \nu_q)^{\top}$ is the vector of parameters of interest and $\tau=(\tau_1, \ldots, \tau_s)^{\top}$ is the vector of nuisance parameters, where $m+p+M=q+s$. Suppose the interest is in testing the null hypothesis $\mathcal{H}_0: \nu = \nu_0$, where $\nu_0$ is a specified vector of constants of size $q$. The likelihood ratio statistic is given by: 
\begin{align*} 
{{\rm{LR}}}=2 \left[ \ell(\widehat{\theta})-\ell(\widetilde{\theta}) \right], 
\end{align*} 
where $\ell(\theta) $ is the log-likelihood function given in Equation~\eqref{E:logverossimilhanca}, evaluated at $\theta=(\nu^{\top}, \tau^{\top})^{\top}$, $\widehat{\theta}=(\widehat{\nu}^{\top}, \widehat{\tau}^{\top})^{\top}$ is the unrestricted MLE of $\theta$, $\widetilde{\theta}=({\nu_0}^{\top}, \widetilde{\tau}^{\top})^{\top}$ is the restricted MLE of $\theta$ (under the null hypothesis). 

Under usual regularity conditions and under $\mathcal{H}_0$, the ${{{\rm{LR}}}}$ statistic has approximately a distribution $\chi^2_q$ with error of order $n^{-1}$~\citep{Casella2002, Pereira2012, Bayer2013}, where $q$ is the number of parameters tested in the null hypothesis. However, in samples of finite size these approximations can be poor, resulting in size distortions. In this context, analytical or numerical/computational adjustments may be considered for inferential improvements in small samples. Following the bootstrap Bartlett correction proposed in this paper for the likelihood ratio statistic in the inflated beta regression model is presented, as well as the Skovgaard adjustment for inflated beta model given in~\cite{Pereira2012}. 

\subsection{Bootstrap Bartlett correction} 

In order to improve the performance of the likelihood ratio test in small samples, in \cite{Bartlett1937} is introduced the Bartlett correction, later generalized by \cite{Lawley1956}. The Bartlett correction is given by: 
\begin{align*} 
{{{\rm{LR}}}}_{\text{Bartlett}}=\dfrac{{{{\rm{LR}}}}}{c}, 
\end{align*} 
where $c = E({{{\rm{LR}}}})/q$ is known as the Bartlett correction factor. The determination of $c$ using Lawley's (\citeyear{Lawley1956})
 notation involves the product of cumulants and mixed cumulants up to fourth order that are not invariant by permutation~\citep{gauss1993}. In beta regression models the analytical obtaining of $c$ can be costly or even impossible, especially for the non orthogonality of parameters \citep{Pinheiro2011, Bayer2013}. For the inflated beta regression model with variable dispersion, considered in this work, the analytical derivation of the Bartlett correction becomes practically intractable. 

As an numerical alternative to analytical derivation of the Bartlett correction, \cite{Rocke1989} introduces the bootstrap Bartlett correction, where the correction factor $c$ is determined via the bootstrap method~\citep{Efron1979}. The bootstrap Bartlett correction becomes a viable alternative to inferential improvements in small samples when there are impeditive or too costly analytical difficulties, as in the model considered here. 

The bootstrap Bartlett correction considering the expected value of ${{{\rm{LR}}}}$, directly estimated from the observed sample $ y = (y_1, \ldots , y_n) ^ {T} $ using bootstrap, can be described by the following steps:
\begin{enumerate} 
\item Generate, under $\mathcal{H}_0$, $B$ bootstrap resamples $(y^{*1}, \ldots , y^{*B})$ of the model, replacing the model parameters by the estimates in $\mathcal{H}_0$ using the original sample (parametric bootstrap). 
\item Obtain the bootstrap ${{{\rm{LR}}}}$ statistic for each pseudosample $y^{*b}$, with $b=1, \ldots, B$, calculated in the following way: 
\begin{align*} 
{{{\rm{LR}}}}^{*b}=2\lbrace \ell(\widehat{\theta}^{*b};y^{*b}) - \ell(\widetilde{\theta}^{*b};y^{*b}) \rbrace, 
\end{align*} 
in which $\widehat{\theta}^{*b}$ is the MLE of $\theta$ under the alternative hypothesis $\mathcal{H}_1$, e $\widetilde{\theta}^{*b}$ is the MLE under $\mathcal{H}_0$. 
\item Calculate the corrected ${{{\rm{LR}}}}$ statistic, given by: 
\begin{align}\label{E:bartlett} {{{\rm{LR}}}}_{B}=\dfrac{{{{\rm{LR}}}}q}{\overline{{{{\rm{LR}}}}^*}}, 
\end{align}
 in which $\overline{{{{\rm{LR}}}}^*}=\dfrac{1}{B}\sum\limits_{b=1}^{B}{{{\rm{LR}}}}^{*b}$. 
\end{enumerate} 

In the bootstrap Bartlett correction the ${{{\rm{LR}}}}$ statistic is corrected so its distribution in small samples can be better approximated by the reference null distribution, $\chi_q^2$ \citep{Bayer2013}. Meanwhile, the usual bootstrap correction consists of obtaining a bootstrap approximation for the null distribution of the test statistic \citep{Queiroz2014}. \cite{Rocke1989} states that the bootstrap Bartlett correction has computational advantages compared to the usual bootstrap scheme, and with $B=100$, in general, there are results equivalent to the usual bootstrap method with $B=700$. Also, through simulation studies, \cite{Bayer2013} conclude that $B$ values larger than $200$ lead to negligible improvements for bootstrap Bartlett correction. In this sense, the bootstrap Bartlett correction has good computational advantages over the usual bootstrap method for hypothesis testing correction. 

\subsection{Skovgaard adjustment} 

Another possible correction of the likelihood ratio statistic is the Skovgaard's adjustment, originally presented in \cite{skovgard1996} and subsequently generalized in \cite{skovgard2001}. This adjustment, obtained analytically, is considerable simpler than the Bartlett correction \citep{Pereira2012}. The Skovgaard's adjustment only require first- and second- order log-likelihood cumulants and, different from the Bartlett correction, independent of the orthogonality of the parameters. 

Skovgaard's approximation has been used in different models. Among them, in the non-linear models of exponential family \citep{cysneiros2008} and in the extreme values models~\citep{Ferrari2014}. In the class of beta regression models we have the Skovgaard adjustment for beta regression model with varying dispersion \citep{Pinheiro2011} and in the inflated beta regression model with varying dispersion \citep{Pereira2012}. The results of these studies indicate that the test based on the Skovgaard statistic performs better than the test based on the uncorrected ${{{\rm{LR}}}}$ statistic. 

The likelihood ratio statistic modified by Skovgaard \cite{skovgard2001} is given by:
\begin{align*} 
 {{{\rm{LR}}}}_{Sk_1}={{{\rm{LR}}}} \left( 1- \dfrac{1}{{{{\rm{LR}}}}}\log \overline{\xi} \right),
\end{align*} 
 in which
 \begin{align*} \overline{\xi}=\vert \tilde{I}\vert ^{1/2}\vert\hat{I}\vert^{1/2}\vert\hat{\Upsilon}\vert^{-1}\vert \tilde{J}_{\tau \tau}\vert^{1/2}\vert [ \tilde{I}\hat{\Upsilon}^{-1} \hat{J} \hat{I}^{-1}\hat{\Upsilon} ]_{\tau\tau} \vert ^{-1/2} \dfrac{\lbrace \tilde{U}^{\top} \hat{\Upsilon}^{-1} \hat{I}\hat{J}^{-1}\tilde{U}\rbrace^{q/2}}{{{{\rm{LR}}}}^{q/2-1} \tilde{U}^{\top}\hat{\Upsilon}^{-1}\hat{r}}, 
 \end{align*} 
 where $I$ is the expected information matrix, $J$ is the observed information matrix, $U$ is the total score function,  $\hat{\Upsilon}=\mathbb{E}_{\hat{\theta}}[U(\hat{\theta})U^{\top}(\tilde{\theta})]$,   $\hat{r}=\mathbb{E}_{\hat{\theta}}[U(\hat{\theta})(\ell(\hat{\theta})-\ell(\tilde{\theta}))]$ and $J_{\tau \tau}$ is the observed information matrix $s \times s$ corresponding to the vector $\tau$. Yet, ``hat'' denotes evaluation in the unrestricted MLE and ``tilde'' the evaluation in the restricted MLE.

An asymptotically equivalent version to ${{{\rm{LR}}}}_{Sk_1}$ is given by: \begin{align*} {{{\rm{LR}}}}_{Sk_2}= {{{\rm{LR}}}} - 2 \log \overline{\xi}. \end{align*} 

Under the null hypothesis, the statistics ${{{\rm{LR}}}}_{Sk_1}$ and ${{{\rm{LR}}}}_{Sk_2}$ have approximately the distribution $\chi^2_q$ with high precision~\citep{Pereira2012}. For details on the analytical derivation of the Skovgaard adjustment in inflated beta regression model, see \cite{Pereira2012}. 

\section{Numerical results}\label{S:resultados} 

To evaluate the performance in small samples of  
the proposed statistic ${{{\rm{LR}}}}_B$, given in \eqref{E:bartlett}, 
the usual likelihood ratio statistic (${{{\rm{LR}}}}$) 
and 
the two versions of the Skovgaard adjustment (${{{\rm{LR}}}}_{Sk_1}$ and ${{{\rm{LR}}}}_{Sk_2}$), 
a simulation study was performed. The number of Monte Carlo replications was 5000 and for the bootstrap Bartlett correction were considered $B=200$ bootstrap resamples. The sample sizes used were $30, \, 40, \, 50$. The entire computational implementation was developed in the language {\tt R} \citep{R2012}, and for the estimation of the model parameters the package {\tt gamlss} \citep{Stasinopoulos2007} was used. 

\begin{table}[h]
\small
 \caption{Null rejection rates $(\%)$; submodels for $\mu$, $\phi$ and $\alpha$} 
\begin{center}
\begin{tabular}{llcrrrrrrrrrr}
\toprule
& & \multicolumn{3}{c}{$1\%$} & \multicolumn{3}{c}{$5\%$}&\multicolumn{3}{c}{$10\%$}  \\
 \cmidrule(r){3-5}  
 \cmidrule(r){6-8}  
 \cmidrule{9-11}
  \vspace{-0.2cm}
 \\
 %\colrule
$q$ & Stat
\backslashbox[9pt][l]{}{}%{Estatística}{$n$} 
$n$  
& $30$ & $40$ & $50$ &  $30$ &  $40$ & $50$ & $30$ & $40$ & $50$\\
\hline
\multicolumn{11}{c}{Submodel for $\mu$}\\
\hline
1& ${{{\rm{LR}}}}$ & $3.16$ &  $2.20$& $2.02$ & $9.84$ & $7.94$ & $7.26 $ & $16.84$ & $13.62$ & $13.66$ \\
& ${{{\rm{LR}}}}_{B}$   & $0.80$ &  $\textbf{0.90}$ & $\textbf{1.18}$ & $4.80$ & $4.84$  & $5.34 $ & $9.22$ & $9.52$ & $10.20$\\ 
& ${{{\rm{LR}}}}_{Sk1}$ &   $\textbf{1.10}$ & $1.18$& $1.48$ & $\textbf{5.08}$ & $5.26$ & $5.22 $ & $\textbf{9.98}$& $10.42$ & $10.50$\\ 
\vspace{0.2cm}
& ${{{\rm{LR}}}}_{Sk2}$ &   $0.76$ & $0.88$ & $1.22$ & $4.50$& $\textbf{4.94}$ &$\textbf{4.92}$ & $8.90$ & $\textbf{9.82}$ & $\textbf{10.18}$ \\
%\colrule
2& ${{{\rm{LR}}}}$ & $3.22$ &  $2.24$ & $2.10$ & $10.06$ & $8.24$ & $7.46$ & $17.22$ & $15.06$ & $13.18$ \\
& ${{{\rm{LR}}}}_{B}$  & $\textbf{0.80}$ &  $\textbf{0.88}$ & $\textbf{1.12}$ & $\textbf{4.96}$ & $\textbf{4.56}$ & $\textbf{4.70}$ & $\textbf{9.50}$ & $\textbf{9.58}$ & $\textbf{9.56}$ \\ 
 & ${{{\rm{LR}}}}_{Sk1}$ & $1.44$ & $1.36$ & $1.40 $ & $6.10$ & $5.68$ & $5.50 $ & $11.64$ & $11.42$ & $10.60$ \\ 
& ${{{\rm{LR}}}}_{Sk2}$ & $1.36$ & $1.36$ & $1.40 $ & $5.90$& $5.60$ & $5.48$ & $11.20$ & $11.28$ & $10.54$\\
\hline
\multicolumn{11}{c}{Submodel for $\phi$}\\
\hline
$1$ & ${{{\rm{LR}}}}$ & $2.54$ & $1.90$ & $1.34$ & $8.34$ & $7.56$ & $6.36$ & $14.80$ & $13.80$& $11.72$  \\
& ${{{\rm{LR}}}}_{B}$ & $0.46$ &  $\textbf{0.90}$  & $0.68$ & $\textbf{3.80}$ & $\textbf{4.58}$ & $4.22$ & $\textbf{8.04}$ & $\textbf{9.64}$ & $8.84$\\ 
& ${{{\rm{LR}}}}_{Sk1}$ & $1.82$ & $1.58$ & $1.44$ & $7.00$ & $6.34$  & $5.94$ & $12.84$& $11.88$ & $11.28$\\ 
\vspace{0.2cm}
& ${{{\rm{LR}}}}_{Sk2}$ & $\textbf{1.32}$ & $1.24$ & $\textbf{1.08}$ & $6.34$& $5.88$ &$\textbf{5.54}$ & $12.00$& $11.18$ & $\textbf{10.74}$ \\
%\colrule
$2$ & ${{{\rm{LR}}}}$ & $2.62$ & $2.24$ & $1.92$ & $9.68$ & $8.78$ & $7.50$ & $16.86$ & $14.62$& $13.66$  \\
& ${{{\rm{LR}}}}_{B}$ & $\textbf{1.00}$ &  $\textbf{0.94}$  & $\textbf{1.10}$ & $\textbf{4.62}$ & $\textbf{5.38}$ & $\textbf{5.08}$ & $\textbf{9.74}$ & $\textbf{10.20}$ & $\textbf{10.42}$\\ 
& ${{{\rm{LR}}}}_{Sk1}$ & $1.54$ & $1.22$ & $1.16$ & $6.74$ & $6.20$  & $5.58$ & $12.80$& $11.24$ & $11.22$\\ 
& ${{{\rm{LR}}}}_{Sk2}$ & $1.40$ & $1.20$ & $1.14$ & $6.30$& $6.04$ &$5.56$ &$12.44$& $11.00$ & $11.08$ \\
\hline
\multicolumn{11}{c}{Submodel  for $\alpha$}\\
\hline
$1$ & ${{{\rm{LR}}}}$ & $1.70$ & $1.70$ & $1.28$ & $6.50$ & $6.38$ & $5.64$ & $12.18$ & $11.88$& $11.12$  \\
& ${{{\rm{LR}}}}_{B}$ & $0.80$ &  $\textbf{1.06}$  & $\textbf{1.04}$ & $4.42$ & $4.86$ & $4.72 $ & $9.18$ & $9.12$ & $9.96$\\ 
& ${{{\rm{LR}}}}_{Sk1}$ & $\textbf{0.84}$ & $1.12$ & $1.06$ & $\textbf{4.68}$ & $5.10$  & $\textbf{4.86}$ & $\textbf{9.76}$& $\textbf{9.76}$ & $\textbf{10.00}$\\ 
\vspace{0.2cm}
& ${{{\rm{LR}}}}_{Sk2}$ & $0.82$ & $1.08$ & $\textbf{1.04}$ & $4.50$& $\textbf{4.98}$ &$4.80$ &$9.46$& $9.62$ & $9.92$ \\
%\colrule
$2$ & ${{{\rm{LR}}}}$ & $1.96$ &  $2.04$ & $1.90$ & $8.46$ & $7.80$ & $6.62$ & $14.22$ & $14.02$& $11.70$  \\
& ${{{\rm{LR}}}}_{B}$ & $0.62$ & $0.80$  & $1.16$ & $3.60$ & $4.80$ & $4.52$ & $8.06$ & $9.54$ & $8.90$\\ 
& ${{{\rm{LR}}}}_{Sk1}$ & $\textbf{1.34}$ & $\textbf{0.96}$ & $1.12$ &$\textbf{4.80}$ & $5.28$  & $\textbf{4.78}$ & $\textbf{9.60}$& $10.46$ & $\textbf{9.54}$\\ 
& ${{{\rm{LR}}}}_{Sk2}$ &  $0.58$ & $0.76$ & $\textbf{1.08}$ & $3.90$& $\textbf{4.88}$ &$4.68$ &$8.64$& $\textbf{10.04}$ & $9.50$ \\
\hline
\end{tabular}
\end{center}
\label{T:tamanho} \end{table} 

All results for evaluating the null rejection rate (size) of the tests are shown in Table~\ref{T:tamanho}, considered the one-inflated beta regression model. In this table the best results are highlighted. Nominal levels were considered equal to $1\%$, $5\%$ and $10\%$. In the evaluation of the tests on the parameters of the mean submodel, it was considered the following regression structure for the mean, precision and mixture parameters : \begin{align*} g(\mu_t)&=\beta_0+\beta_1 x_{1t}+\beta_2 x_{2t},\\ b(\phi_t)&=\lambda_0+\lambda_1 s_{1t},\\ h(\alpha_t)&=\gamma_0+\gamma_1 z_{1t}, \end{align*} in which $t=1,\ldots,n$. For the structure of mean regression, $g(\mu_t)$, and mixture, $h(\alpha_t)$, the logit link function was used and for the structure of precision parameter, $b(\phi_t)$, the logarithmic link function. 

In the Monte Carlo simulation, we consider two scenarios for the null hypothesis: (i) $q=1$, in which $\mathcal{H}_0: \beta_2=0$, fixing the parameters $\beta_0=-1$, $\beta_1=3.5$, $\beta_2=0$, $\lambda_0=5.1$, $\lambda_1=-2.8$, $\gamma_0=-2$, $\gamma_1=1.5$; and (ii) $q=2$, $\mathcal{H}_0:\beta_1= \beta_2=0$, where $\beta_0=2$, $\beta_1=\beta_2=0$, with the same parameter values for $\phi$ and $\alpha$ submodels considered for $q=1$. 
These values for the parameters in (i) imply the averages of $y$ and $\phi$ to be equal, respectively, to $0.731$ and $55.102$, when $n=50$. 
For (ii), the averages of $y$ and $\phi$ are, respectively, equal to $0.908$ and $55.102$, with $n=50$. The matrix of regressors is generated from a standard uniform distribution, $\mathcal{U}(0,1)$, and kept constant during all Monte Carlo replications. For each replication, a sample $y_1 ,\ldots,y_n$ is generated with one-inflated beta distribution given by \eqref{E:densidadeinflacionada}. 

We also consider tests on the parameters of the submodel for precision ($\phi$). In these cases we consider the one-inflated beta regression model given by: 
\begin{align*} 
g(\mu_t)&=\beta_0+\beta_1 x_{1t},\\ 
b(\phi_t)&=\lambda_0+\lambda_1 s_{1t}+\lambda_2s_{2t},\\ 
h(\alpha_t)&=\gamma_0+\gamma_1 z_{1t}. 
\end{align*} 
To evaluate the null rejection rate of the tests, it was considered the following scenarios: (i) $q=1$, $\mathcal{H}_0: \lambda_2=0$, fixing the parameters $\beta_0=-1$, $\beta_1=3.5$, $\lambda_0=5.1$, $\lambda_1=-2.8$, $\lambda_2=0$, $\gamma_0=-2$, $\gamma_1=1.5$; and (ii) $q=2$, $\mathcal{H}_0:\lambda_1= \lambda_2=0$, considering $\lambda_0=5.1$, $\lambda_1=\lambda_2=0$. 
The average values of $y$ and $\phi$ in this scenario are, respectively, equal to $0.728$ and $54.865$, for (i) with $n=50$.
For (ii), with $n=50$, the averages of $y$ and $\phi$ are, respectively, equal to $0.728$ and $164.022$.

Further, to evaluate the null rejection rate of the tests to make inferences about the parameters of the $\alpha$ submodel, we considered the following regression structure: 
\begin{align*} 
g(\mu_t)&=\beta_0+\beta_1 x_{1t},\\ 
b(\phi_t)&=\lambda_0+\lambda_1 s_{1t},\\ 
h(\alpha_t)&=\gamma_0+\gamma_1 z_{1t}+\gamma_2 z_{2t}. 
\end{align*} 
In this case, were considered: (i) $q=1$, $\mathcal{H}_0: \gamma_2=0$, fixing the parameters $\beta_0=-1$, $\beta_1=3.5$, $\lambda_0=5.1$, $\lambda_1=-2.8$, $\gamma_0=-2$, $\gamma_1=1.5$; and (ii) $q=2$, $\mathcal{H}_0:\gamma_1= \gamma_2=0$,  considering $\gamma_0=-2$. 
These values for the parameters in (i) imply averages of $y$ and $\phi$ equal, respectively, to $0.728$ and $55.001$, when $n=50$. 
For (ii), the averages of $y$ and $\phi$ are, respectively, equal to $0.688$ and $55.001$, with $n=50$.

Examining the Table \ref{T:tamanho}, where are presented the results of tests' size, considering the $\mu$ submodel, it is found that the ${{{\rm{LR}}}}$ test is the most liberal, showing rejection rates well above nominal levels. 
For example, at the level of $5\%$ and $10\%$ for $n=30$ and $q=2$, 
the rejection rates for ${{{\rm{LR}}}}$ are, respectively, $10.06\%$ and $17.22\%$. 
The corrected statistics, both the bootstrap Bartlett correction as well as the two versions of Skovgaard adjustment, have less size distortion than the test considering the usual uncorrected statistical. When imposed only one restriction, i. e., $q=1$, the ${{{\rm{LR}}}}_{B}$ showed good performance, but the ${{{\rm{LR}}}}_{Sk1}$ statistic showed the best results for $n=30$.
For $q=2$, the proposed ${{{\rm{LR}}}}_{B}$ statistic has the best performance in all sample sizes and significance levels. Still, among the corrected statistics, the more liberal is ${{{\rm{LR}}}}_{Sk1}$, i. e., it has in general higher rejection rate than the nominal level. For this liberal characteristic of ${{{\rm{LR}}}}_{Sk1}$, it is already expected that its results on the evaluation of tests' power will be higher. 

For the results of tests' size on the submodel parameters of $\phi$ it can also be verified that the corrected statistics have better results. In particular, we highlight the performance of the proposed statistic ${{{\rm{LR}}}}_B$ when imposed two restrictions on the null hypothesis. Also, it can be seen that the versions corrected by Skovgaard are more liberal. 
For example, at the level of $10\%$   
the null rejection rates of the ${{{\rm{LR}}}}_{Sk1}$ are $12.80\%$ ($n=30$), $11.24\%$ ($n=40$) and $11.22\%$ ($n=50$).

\begin{table}%[t]
\small
\caption{Estimated quantiles and moments of the test statistics for the submodel for $\mu$, $q=2$ and $n=40$}
\begin{center}
\begin{tabular}{lclllllllll}
\hline
Variate & Mean & Variance & Skewness & Kurtosis & $90$th-perc & $95$th-perc & $99$th-perc  \\
\hline
$\chi^2_q$ & $2.000$ & $4.000$ & $2.000$ & $9.000$ &  $4.605$ & $5.991$  & $9.210$ \\
 ${{{\rm{LR}}}}$ & $2.400$ &  $5.438$  &  $1.828$ & $7.345$  & $5.526$ &$7.076$ & $11.015$\\
${{{\rm{LR}}}}_B$  & $\textbf{1.963}$ & $\textbf{4.028}$ & $\textbf{1.831}$ & $\textbf{7.365}$  & $\textbf{4.558}$ & $\textbf{5.827}$ &  $\textbf{9.000}$ \\ 
 ${{{\rm{LR}}}}_{Sk1}$ & $2.105$ & $4.456$ & $1.819$ & $7.262$  & $4.884$& $6.236$ &  $9.604$\\  
 ${{{\rm{LR}}}}_{Sk2}$ & $2.089$ & $4.431$  & $1.818$ & $7.260$ & $4.859$ & $6.213$ &  $9.592$\\
 \hline
\end{tabular}
\end{center}
\label{T:momentos}
\end{table}

\begin{figure}%[t]
\subfigure[$n=30$.]{\includegraphics[width=0.325\textwidth]{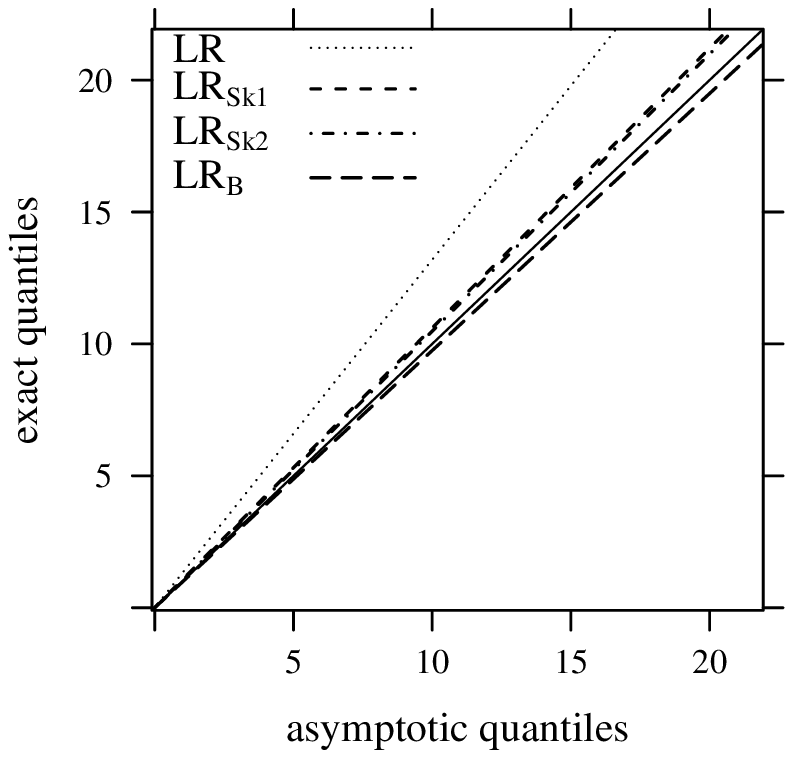}\label{f:res30}}
\subfigure[$n=40$.]{\includegraphics[width=0.325\textwidth]{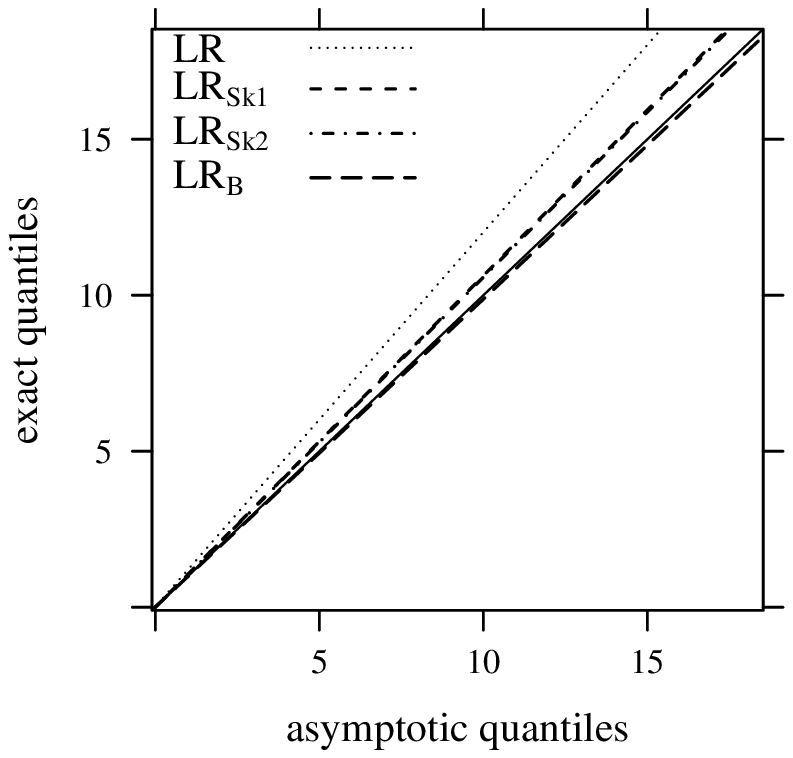}\label{f:res40}}
\subfigure[$n=50$.]{\includegraphics[width=0.325\textwidth]{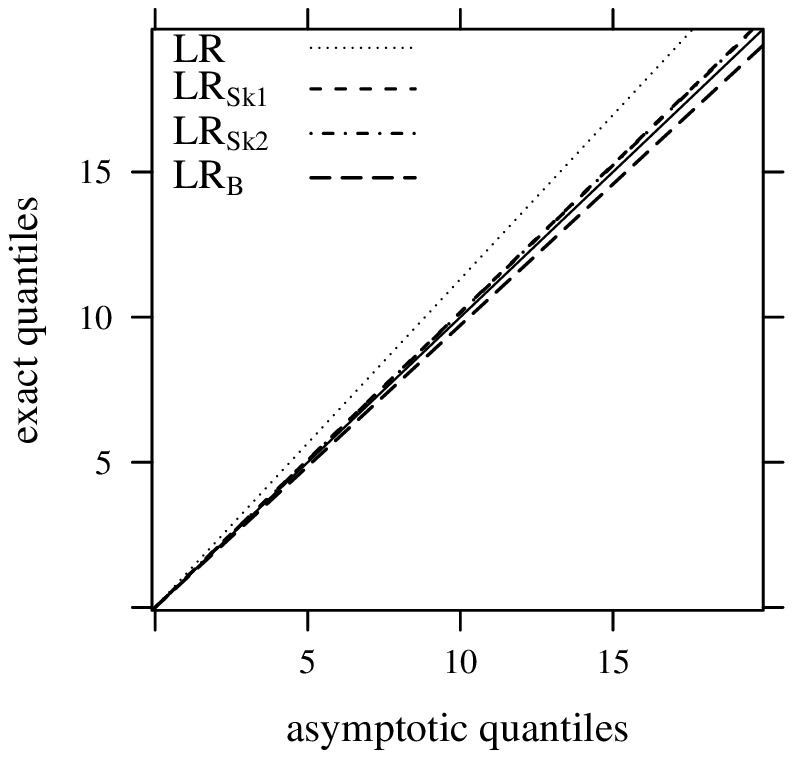}\label{f:res50}}
\caption{Quantile-Quantile graph for the submodel of $\mu$, $q=2$ and different sample sizes.}\label{F:qqplots}
\end{figure}

\begin{table}[!h]
\small
\caption{Non-null rejection rates $(\%)$, for the submodels for $\mu$, $\phi$ and $\alpha$, subject to a restriction ($q=1$)}
\label{T:poder}
\begin{center}
\begin{tabular}{clrrrrrrrrrrr}
\hline
& & \multicolumn{2}{c}{$1\%$}&& \multicolumn{2}{c}{$5\%$}&&\multicolumn{2}{c}{$10\%$}  \\
 \cmidrule(r){3-4}  
 \cmidrule(r){6-7}  
 \cmidrule{9-10}
  \vspace{-0.2cm}
 \\
%\colrule
$\delta$ & Stat \backslashbox[10pt][l]{}{} $n$ & $30$ & $50$ &&  $30$ & $50$ && $30$ & $50$\\
\hline
&\multicolumn{8}{c}{Submodel for $\mu$}\\
\hline
$-1$ & ${{{\rm{LR}}}}_{B}$   & $84.92$  & $96.62$ & & $96.68$ & $99.56$ && $98.40$ & $99.82$\\ 
& ${{{\rm{LR}}}}_{Sk1}$ &   $87.06$ & $97.06$ & &$97.02 $ & $99.60$ && $98.48$& $99.86$\\ 
\vspace{0.1cm}
& ${{{\rm{LR}}}}_{Sk2}$ &   $85.62$  & $96.96$ & & $96.48$ &$99.60$ && $98.36$ & $99.84$ \\
$-0.5$ & ${{{\rm{LR}}}}_{B}$   & $28.18$  & $38.36$ & & $56.02$ & $64.20$ && $69.56$ & $75.68$\\ 
& ${{{\rm{LR}}}}_{Sk1}$ &   $30.36$ & $39.08$ & &$57.84$  & $65.06$ && $70.88$& $75.76$\\ 
\vspace{0.1cm}
& ${{{\rm{LR}}}}_{Sk2}$ &   $28.48$ & $38.76 $ & & $56.12$  &$64.80$ && $69.48$ & $75.44$ \\
$0.5$ & ${{{\rm{LR}}}}_{B}$   & $23.80$  & $38.10$ & & $50.56$& $64.04$  && $65.04$ & $75.22$\\ 
& ${{{\rm{LR}}}}_{Sk1}$ &   $26.50$ & $39.32$ & &$52.78$ & $64.56$ && $ 70.88$& $75.78$\\ 
\vspace{0.1cm}
& ${{{\rm{LR}}}}_{Sk2}$ &   $25.04$ & $38.92$ & & $51.12$ & $64.22$ && $64.88$ & $75.56$ \\
$1$ & ${{{\rm{LR}}}}_{B}$ & $78.12$  & $92.90$ & & $93.48$ &$98.28$ && $96.90$ & $99.32$\\ 
& ${{{\rm{LR}}}}_{Sk1}$ & $80.44$ & $93.44$ & &$94.08$  & $98.56$ && $97.12$&  $99.38$\\ 
& ${{{\rm{LR}}}}_{Sk2}$ & $79.18$ & $93.24$ & & $93.54$  &$98.48$ && $96.88$  & $99.38$ \\
\hline
&\multicolumn{8}{c}{Submodel for $\phi$}\\
\hline
$-4$ & ${{{\rm{LR}}}}_{B}$   &    $82.64$  & $97.70$ && $92.98$ & $ 99.50$ && $96.16$ & $99.76$\\ 
& ${{{\rm{LR}}}}_{Sk1}$ &   $88.68$  & $98.18$ && $96.02$ & $ 99.58$ && $97.96$ & $99.84$\\
\vspace{0.1cm}
& ${{{\rm{LR}}}}_{Sk2}$ &  $88.58$  & $98.18$ && $95.80$ & $ 99.58$ && $97.86$ & $99.84$\\
$-3$ & ${{{\rm{LR}}}}_{B}$ & $56.06$  & $81.30$ && $77.10$ & $92.50$ && $84.34$ & $95.54$\\ 
& ${{{\rm{LR}}}}_{Sk1}$ &   $65.74$  & $83.24$ && $82.48$ & $93.08$ && $88.82$ & $96.00$\\
\vspace{0.1cm}
& ${{{\rm{LR}}}}_{Sk2}$ &  $65.44$  & $83.14$ && $82.24$ & $93.04$ && $88.62$ & $96.00$\\
$3$ & ${{{\rm{LR}}}}_{B}$ & $46.02$ & $ 70.24$ && $70.40$ & $86.92$ && $80.70 $ & $92.22$\\ 
& ${{{\rm{LR}}}}_{Sk1}$ & $48.68$ & $72.00$ && $71.94$ & $87.94 $ && $81.68 $ & $92.94$\\
\vspace{0.1cm}
& ${{{\rm{LR}}}}_{Sk2}$ & $47.64 $ & $71.90$ && $70.48$ & $87.78$ && $80.76 $ & $92.80$\\
$4$ & ${{{\rm{LR}}}}_{B}$  & $ 72.58$  & $92.76$ && $88.44$ & $97.70$ && $92.82$ & $99.10$\\ 
& ${{{\rm{LR}}}}_{Sk1}$ &   $74.60$  & $93.74$ && $88.62$ & $98.00$ && $92.86$ & $99.10$\\
& ${{{\rm{LR}}}}_{Sk2}$ &  $73.58$  & $93.64$ && $87.86$ & $98.00$ && $92.10$ & $99.10$\\
\hline
&\multicolumn{8}{c}{Submodel for $\alpha$}\\
\hline
$1$ & ${{{\rm{LR}}}}_{B}$ & $2.40$  & $5.84  $ && $8.62$ & $  17.28$ && $15.10$ & $26.33 $\\ 
& ${{{\rm{LR}}}}_{Sk1}$ & $ 2.58$ & $ 5.84$ && $8.64$ & $17.34 $ && $15.66$ & $26.10 $\\
\vspace{0.1cm}
& ${{{\rm{LR}}}}_{Sk2}$ & $2.42 $ & $5.76 $ && $ 8.42$ & $17.22 $ && $ 15.48$ & $26.02 $\\
$2$ & ${{{\rm{LR}}}}_{B}$  & $8.72$  & $30.22$ && $23.54$ & $54.64 $ && $35.16$ & $67.02 $\\ 
& ${{{\rm{LR}}}}_{Sk1}$ &   $8.72$  & $30.18$ && $23.10$ & $55.14 $ && $35.28$ & $66.88$\\
& ${{{\rm{LR}}}}_{Sk2}$ &  $8.58$  & $30.10$ && $22.78$ & $55.12 $ && $35.04$ & $66.84 $\\
\hline
\end{tabular}
\end{center}
\end{table}

For inferences about the submodel parameters of $\alpha$, as shown in Table \ref{T:tamanho}, the best results are also shown by the corrected statistics. As expected, tests on the parameters that index the mixture parameter submodel have very similar results to results for inferences about the regression structures $\mu$ and $\phi$. In general, the Skovgaard adjustments show better performance in this case, however, the ${{{\rm{LR}}}}_B$ statistic still has similar and much higher performance than the usual likelihood ratio. 

The objective of the second order corrections considered here is to improve the approximation of the ${{{\rm{LR}}}}$ test statistic distribution by the null chi-squared limit distribution. Table~\ref{T:momentos} presents quantiles and estimated moments of the considered statistics, as well as the reference values of $\chi^2_q$. The scenario testing the submodel parameters of $\mu$, under two restrictions, $q = 2$, and with $n=40$ was considered for these results. It is verified that the statistic distribution of ${{{\rm{LR}}}}$ is the farthest from reference chi-squared distribution. Among the four statistics considered, those having moments and quantiles closer to $\chi_q^2$ is the proposed ${{{\rm{LR}}}}_B$. Still, it is observed that in general the corrected statistics present values of calculated measures closer to the reference values of $\chi_q^2$ than the ${{{\rm{LR}}}}$.

Figure~\ref{F:qqplots} shows the QQ-plot graphs (exact quantiles versus asymptotic quantiles) for different sample sizes, given the same scenario of the results of Table~\ref{T:momentos}. It's clear that the distribution of the proposed statistic is much closer to the reference null distribution, $\chi_q^2$. It was also observed that all the corrected statistics are closer to the reference null distribution of the usual ${{{\rm{LR}}}}$ statistic.

Table \ref{T:poder} presents the results of Monte Carlo simulations for non-null rejection rate (power) of the tests on the parameters of the submodels of $\mu$, $\phi$ and $\alpha$. Since the results of simulations of the test size using the ${\rm{LR}}$ statistic are pretty liberal, we present only the results for ${\rm{LR}}_B$, ${\rm{LR}}_{Sk1}$ and ${\rm{LR}}_{Sk2}$. For the mean submodel, we tested $\mathcal{H}_1: \beta_2=\delta$ $(q=1)$, where $\delta=-1, -0.5, 0.5, 1$. For the submodel of $\phi$ we tested $\mathcal{H}_1: \lambda_2=\delta$ $(q=1)$, where $\delta=-4, -3, 3, 4$. Also, about the regression structure of $\alpha$, the tested hypotheses were $\mathcal{H}_1: \gamma_2=\delta$ $(q=1)$, where $\delta=1,2$. 

Based on Table \ref{T:poder} it is noticed that the performances of the three statistics do not differ much for the three submodels. The corrected statistic ${\rm{LR}}_{Sk1}$, in most scenarios, is slightly more powerful. However, this result was expected, for being the most liberal among the corrected statistics. Simulations of power under two constraints ($q=2$) were also considered. However, the results for $q=1$ and $q=2$ are similar and the results for $q=2$ were omitted for briefness. 

Based on the results presented, it is verified the good performance of the bootstrap Bartlett statistic proposed here for inferences in small samples. ${\rm{LR}}_B$ was shown to be equivalent or superior (in some cases) to the Skovgaard analytical adjustment. Whereas the adjusted tests behave more accurately and obtaining the proposed corrected statistic is simpler because it does not require expensive analytical calculations, we recommend using the test based in the bootstrap Bartlett statistic. 

\section{An application}\label{s:aplicacao} 

This section presents an application to real data of the likelihood ratio test corrected via bootstrap Bartlett, proposed in Section~\ref{S:TesteRV}. The data used are part of the work presented in \cite{souza2005} which estimates levels of efficiency for the Brazilian municipalities. These indexes take values in the range $(0,1]$, where $1$ corresponds to the fully efficient municipalities. In this application were considered the $26$  Brazilian state capitals, referent to the $2000$ year. 
The proportion of ones in this data is equal to $0.12$.

The variables considered in the database were: 
number of inhabitants $(x_{1})$, information $(x_{2})$, which is a binary variable that assumes a value of $1$ if the municipality is computerized, and $0$ otherwise, personnel expenses $(x_{3})$, population density $(x_{4})$, percentage of households whose head earns up to $1$ minimum wage $(x_{5})$, urbanization rate $(x_{6})$, index actualization of the real state register $(x_{7})$, a binary variable that receives values $1$ if the municipality is located in areas of the drought polygon area and $0$ otherwise $(x_{8})$ and average income $(x_{9})$. Further details on these and other related variables can be accessed at \cite{souza2005}. 

For the mean submodel, the initial model has been obtained by the function {\tt stepGAIC} of the {\tt gamlss} package available at {\tt R}~\citep{R2012}. This function selects a model by a stepwise algorithm using the generalized Akaike information criteria. For the submodels of $\phi$ and $\alpha$ the same covariates presented in \cite{Pereira2012} were considered. Thus, initially we consider the following model
\begin{align*}
\log\left( \dfrac{\mu_t} {1 -\mu_t} \right) = &\, \beta_0+\beta_1x_{1t}+\beta_2x_{2t}+\beta_3x_{3t}+\beta_4x_{4t},\\ \log(\phi_t) = &\, \lambda_0+\lambda_1x_{9t},\\ \log \left( \dfrac{\alpha_t}{1-\alpha_t} \right) = &\, \gamma_0+\gamma_1x_{9t}. 
\end{align*} 

The tests were performed at the $10\%$ nominal level. When testing the exclusion of the covariate $x_{4}$, $\mathcal{H}_0: \beta_4=0$, we have the values of the statistics and ($p$-value in parenthesis) given by: ${{{\rm{LR}}}}=3.609$ $(p=0.057)$ and ${{{\rm{LR}}}}_B=2.177$ $(p=0.140)$. It is noticed that inferential conclusions using the corrected and non-corrected statistics are opposite. By the corrected ${{{\rm{LR}}}}_B$ statistic, the hypothesis $\mathcal{H}_0$ is not rejected, then we decided to exclude the covariate $x_{4}$ of the submodel. To test the significance of $x_{3}$, $\mathcal{H}_0: \beta_3=0$, we have: ${{{\rm{LR}}}}=5.909$ $(p=0.015)$ and ${{{\rm{LR}}}}_B= 3.837$ $(p=0.050)$; both tests reject the null hypothesis, so $x_{3}$ remains in the submodel. 
When testing $\mathcal{H}_0: \beta_2=0$, it is obtained ${{{\rm{LR}}}}=2.054$ $(p=0.152)$ and ${{{\rm{LR}}}}_B=1.509$ $(p=0.219)$, the null hypothesis is not rejected, then we exclude the covariate $x_{2}$ of the submodel. Yet, for
$\mathcal{H}_0: \beta_1=0$, we have ${{{\rm{LR}}}}=8.287$ $(p=0.004)$ and ${\rm{LR}}_B=6.229$ $(p=0.013)$, in which both reject the null hypothesis. Based on the test corrected via bootstrap Bartlett, the adjusted model is given by: \begin{align*} \log \left( \dfrac{\mu_t}{1-\mu_t} \right) = &\, \beta_0+\beta_1x_{1t}+\beta_3x_{3t},\\ \log(\phi_t)= &\, \lambda_0+\lambda_1x_{9t},\\ \log \left( \dfrac{\alpha_t}{1-\alpha_t} \right) = &\, \gamma_0+\gamma_1x_{9t}. 
\end{align*}

To evaluate the quality of the fitted model, based on the corrected test, we consider the proposed residual analysis in \cite{Ospina2012}. Figure \ref{F:residuosapli} presents the quantile randomized residual graph and the half-normal probability graph with simulated envelope. In Figure \ref{f:tab1}, it is verified that all residual were within the range $(-2, 2)$. Yet, in Figure \ref{f:tab2}, it can be seen that all the points are within the confidence bands of the simulated envelope, indicating a good fit of the model. 

\begin{figure}
\centering
\subfigure[Residuals versus indexes.]{\includegraphics[width=0.4\textwidth]{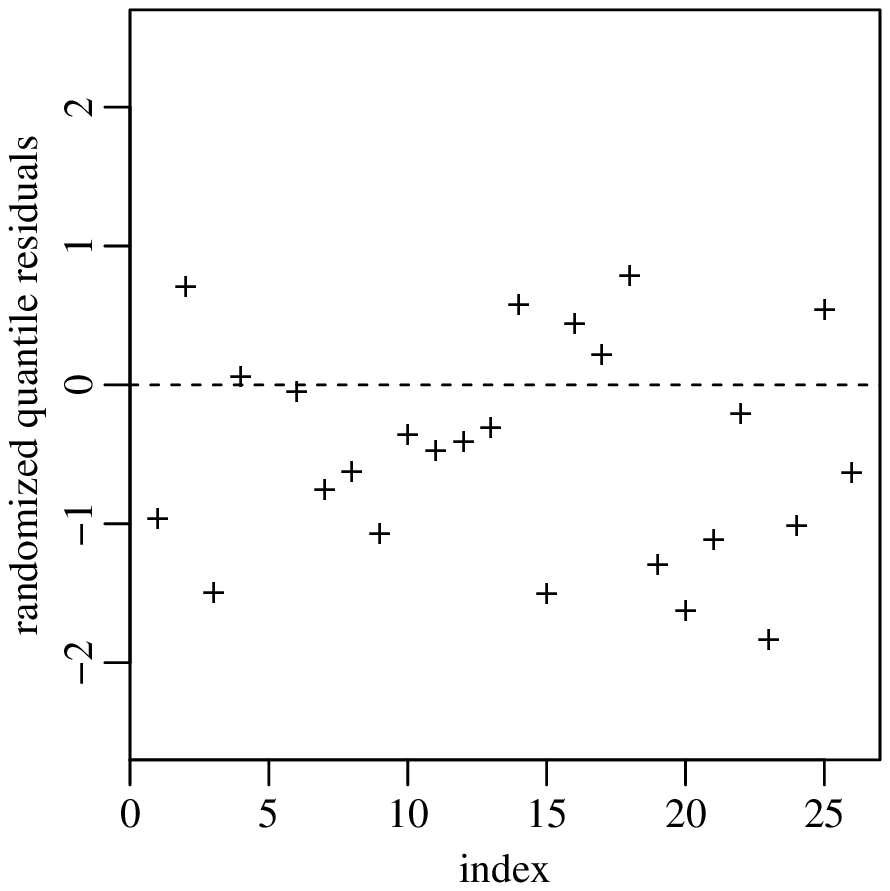}\label{f:tab1}} 
\subfigure[Half-normal probability plot.]{\includegraphics[width=0.4\textwidth]{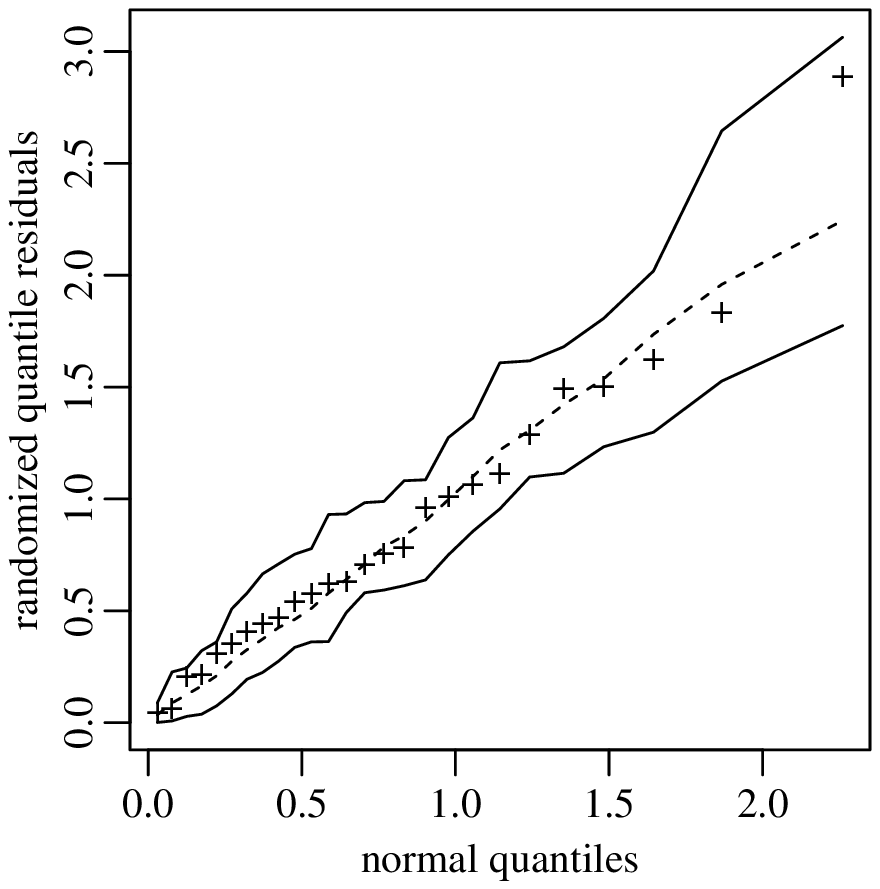}
\label{f:tab2}}
\caption{
Randomized quantile residual plots.
}\label{F:residuosapli}
\end{figure} 

To test whether the model is correctly specified, we consider the RESET test for the inflated beta model presented in \cite{pereirab2014}. In this test we obtained $p=0.997$, not rejecting the null hypothesis that the model is correctly specified. 
 
Therefore, it appears that the model selected based on hypothesis testing using the bootstrap Bartlett corrected test provides a good fit.

\section{Conclusions}\label{s:conclu} 

The likelihood ratio statistic is typically used to perform hypothesis testing in the inflated beta regression models. However, if the sample is not large enough to guarantee a good agreement between the distribution of the test statistic and the limiting $\chi^2$ distribution, the approximate likelihood ratio test can be considerably oversized.
% Inferential improvements in this sense for the inflated beta regression model were presented in \cite{Pereira2012}, where it is proposed the Skovgaard adjustment. 
In this paper we propose a bootstrap Bartlett correction of the likelihood ratio statistic for inferential improvements in the inflated beta regression model in small samples. Through Monte Carlo simulations we evaluated the proposed correction and compared it with the Skovgaard adjustments \citep{Pereira2012} and with the non-corrected usual statistic. The simulation results indicate that the corrected statistics make the tests more accurated, reducing the problem of size distortion in small samples. Still, it is verified that the proposed correction via bootstrap Bartlett has results very close to or even better than the analytical Skovgaard adjustments. The latter requires second-order derivatives of the log-likelihood of the model, while the proposed correction requires only the use of a simple Monte Carlo simulation. 
We believe that the proposed bootstrap Bartlett correction can be quite useful in practical situations and we
recommend to practitioners to model data using inflated beta regressions and use it since it is easy to obtain and present accurate inferential results.

\section*{Acknowledgements}

The authors acknowledge the financial support received by CAPES, FAPERGS and CNPq, Brazil. We also thank two referees for comments and suggestions.

\singlespacing

\bibliography{betareg}

\end{document}